# High-Efficiency Light-Emitting Diodes Based on Formamidinium Lead Bromide Nanocrystals and solution processed transport layers

Francesco Di Stasio[†], Iñigo Ramiro[†,‡], Yu Bi[†,‡], Sotirios Christodoulou[†], Alexandros Stavrinadis[†] and Gerasimos Konstantatos[†,*]

[†]ICFO-Institut de Ciencies Fotoniques, The Barcelona Institute of Science and Technology, 08860 Castelldefels (Barcelona), Spain
[*]ICREA—Institució Catalana de Recerca i Estudis Avançats, Passeig Lluís Companys 23, 08010 Barcelona, Spain

**ABSTRACT:** Perovskite nanocrystal light-emitting diodes (LEDs) employing architecture comprising a ZnO nanoparticles electron-transport layer and a conjugated polymer hole-transport layer have been fabricated. The obtained LEDs demonstrate a maximum external-quantum-efficiency of 6.04%, luminance of 12998 Cd/m$^2$ and stable electroluminescence at 519 nm. Importantly, such high efficiency and brightness have been achieved by employing solution processed transport layers, formamidinium lead bromide nanocrystals (CH(NH$_2$)$_2$PbBr$_3$ NCs) synthesized at room-temperature and in air without the use of a Schlenk line, and a procedure based on atomic layer deposition to insolubilize the NC film. The obtained NCs show a photoluminescence quantum yield of 90% that is retained upon film fabrication. The results show that perovskite NC LEDs can achieve high-performance without the use of transport layers deposited through evaporation in ultra-high-vacuum.

Perovskite nanocrystals (NCs)[1] have been successfully applied in a variety of optoelectronic devices such as solar cells,[2–4] solar concentrators,[5] lasers,[6,7] white phosphors[8] and light-emitting diodes (LEDs)[9–13] Since the synthesis of CsPbX$_3$ NCs (where X = Cl, Br or I) was demonstrated in 2015,[14] attention toward this class of perovskite materials has considerably grown, and more effort has been dedicated in exploiting their favourable optical properties. In particular, focus has been placed on the fabrication of efficient light-emitting diodes (LEDs), since perovskite NCs possess desirable light-emission properties such as tuneable colour and narrow emission, which could lead to colour-pure devices for application in displays and illumination. Another important aspect is that perovskite NCs possess a defect-tolerant structure[1] where defect states do not have a strong influence on the radiative recombination, thus leading to high photoluminescence quantum yield (PLQY) without the need of surface passivation. Nevertheless, major challenges are still present as the labile nature of the surface ligands combined with the strong ionic character of the structure are causes of instability;[15] and purification of the NCs after synthesis requires particular attention as use of anti-solvents can irreversibly damage the material.[4] As a result, their surface chemistry has been studied and tailored[4,7,15] to increase their stability, while purification procedures have been developed to avoid degradation.[4,10] These developments have allowed the fabrication of efficient LEDs and currently the best perovskite NC LEDs show a maximum external-quantum-efficiency (EQE) of 12.9%.[16] Such performance has been achieved in a device structure employing an electron-transport layer prepared trough thermal-evaporation in ultra-high-vacuum conditions. Yet, device architectures based on solution-processed materials, which do not require lengthy and expensive fabrication procedures, are still trailing behind with EQEs in the region of only few percentages.[11,17] The discrepancy in performance has a twofold explanation: first, perovskite NCs possess limited thermal stability which impedes annealing (typically employed to partially insolubilize the layer) without damaging the film.[18] Secondly, anti-solvents (i.e. polar solvents) cannot be used during the spin-coating of top layers as the perovskite NC film will be damage as previously discussed.[4]

Here, we report an approach to overcome these challenges and obtain highly efficient fully-solution processed LEDs based on CH(NH$_2$)$_2$PbBr$_3$ NCs (FAPbBr$_3$). Substituting Cs with formamidinium (FA) in our synthesis, we obtained NCs which are more resistant to the device fabrication procedure. The obtained FAPbBr$_3$ NC film was insolubilized using a procedure based on atomic layer deposition (ALD). The obtained devices show a maximum EQE of 6.04% and a maximum luminance of 12998 Cd/m$^2$.

**Figure 1a** presents a scheme and focused-ion-beam cross-section of the fabricated devices (see Figure S1 in the supporting information for the ultra-violet photoelectron spectroscopy spectra used to derive the FAPbBr$_3$ NC film valence band position with respect to vacuum). As electron-transport layer we have employed a ZnO nanoparticles[19] film deposited via spin-coating on a patterned-ITO/glass substrate, while for hole injection/transport we have used a bi-layer structure comprising poly vinyl carbazole (PVK) and Poly[N,N'-bis(4-butylphenyl)-N,N'-bisphenylbenzidine] (PTPD). Such a hole-transport layer has been previously successfully employed in LEDs based on cadmium chalcogenides NCs as it combines the high hole mobility from the PTPD and the enhanced electron blocking capabilities from the PVK.[20,21] The device was completed via sputtering deposition of a Pt electrode, and encapsulation with epoxy glue and a glass slide. The energy diagram in Figure 1b shows that only small energy barriers are present at the active-layer/transport-layer interface in the LEDs, namely 0.4 eV for electrons and 0.5 eV for holes.

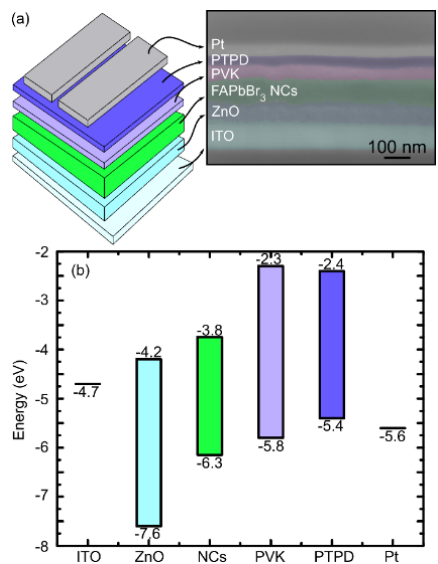

**Figure 1.** (a) Scheme of the fully solution-processed structure used in the light-emitting diodes and FIB SEM cross-sectional image of one device. (b) Energy level diagram of the materials that constitute the LED. The energy values for the various device components were taken from literature, while the values reported for the FAPbBr3 NCs were obtained from ultra-violet photoelectron spectroscopy of a film deposited on ITO, and its optical absorption spectrum.

In **Figure 2a** we report the current density-luminance-voltage (JVL) curve for the "champion device" (i.e. LED with highest EQE). The LED reaches a maximum luminance of 12998 cd/m$^2$; a remarkable value but lower than the 22830 cd/m$^2$ achieved in the highest EQE perovskite LED based on evaporated transport layers,[16] while LEDs based on bulk CsPbBr$_3$ can achieve higher luminance compared to NCs.[22] From the JVL curves we can estimate a turn-on voltage ($V_{on}$, extracted from a luminance of 0.1 cd/m$^2$, see Figure S2) of 4.1 V, which is higher than the estimated optical band-gap of the FAPbBr$_3$ NCs (2.43 eV). In fact, in an ideal LED, the lowest limit of the $V_{on}$ is represented by the optical band gap of the emitting material;[23] yet many factors during device fabrication can induce increased $V_{on}$. In our case, we assign the relatively high $V_{on}$ to the increased thickness of the LED and a corresponding increase in series resistance. We found that using a thick FAPbBr$_3$ NC layer (95 nm) allows the device to reach higher luminance and higher EQE at the cost of an increased operational voltage (see Table S1 in the supporting information). The cross-section of the LED in Figure 1a reveals that the overall device thickness is ≈ 270 nm, thus corroborating our explanation. Despite the high operating voltage, the champion device shows a maximum EQE of 6.04% (Figure 2b) and a maximum current efficiency of 20.53 cd/A. To our knowledge, these are the highest efficiency values reported for perovskite NC LEDs based on solution processed transport layers to date, and the champion device outperforms other LEDs based on FAPbBr$_3$ NCs as well. No drastic efficiency droop is observed for high driving currents: the EQE is reduced only by 10% from 20 to 68.8 mA/cm$^2$ (i.e., 6.3 to 8.4 V).). The efficiency droop in NC LED is known to be caused by Auger recombination at high driving current density when charge balance in the active layer is not achieved. The stable efficiency observed in our LEDs indicates that Auger recombination does not play a major role (i.e. in the active layer charge imbalances are reduced) allowing high luminance to be reached and the efficiency maintained. From the EQE value we can estimate the internal quantum efficiency (IQE) following a previously reported method,[24] thus obtaining a maximum IQE of 25.54%. The estimated IQE is lower than the measured PLQY for the FAPbBr$_3$ NC film used in the device of 46 ± 5%, indicating that further device optimization can increase the efficiency of these LEDs. The electroluminescence (EL) spectrum is stable under increasing applied bias (see Figure 2c, colour coordinates diagram is reported in Figure S3) up to 8.5V. The EL is centred at 519 nm with a full-width-half-maximum (FWHM) of 18 nm, in the range of previously reported perovskite NC LEDs.[10,16] The "champion device" here presented shows good performance in terms of EQE and maximum luminance and this result is reproducible in large part of our LEDs, as shown in the histogram in Figure 2d, where an average EQE of 5.16% is calculated from a total of 30 different devices. Nevertheless, device stability remains very limited as the LEDs are operational for only few minutes, similar to what recently reported by F. Yan et al.[16] The short operational lifetime is currently the main issue of perovskite NC devices and in our case we can observe a decrease in performance already after the initial JVL scan (see Figure S4).

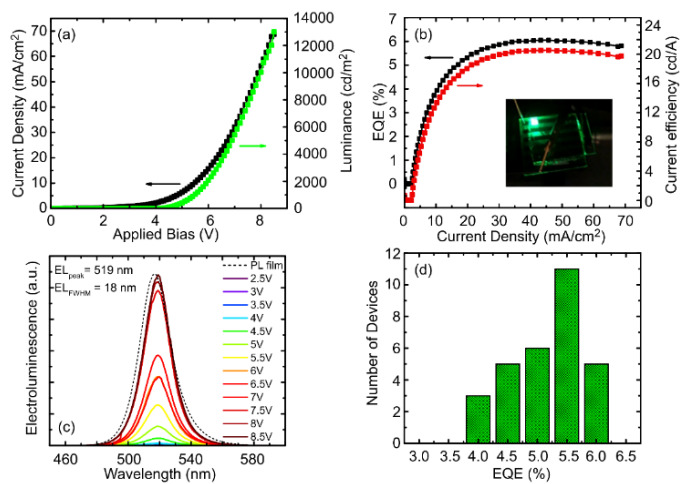

**Figure 2.** (a) Current density-Voltage-Luminance curves for the LED with highest EQE (champion device). (b) EQE and current efficiency curves vs current density, inset: photo of an LED under operation. (c) Electroluminescence spectra recorded at increasing applied voltage. For comparison the PL spectrum of the film is plotted (black dashed line). (d) Histogram of the achieved EQE for the LEDs.

We now focus our discussion on how we achieved such LED performance, starting from the material development and properties, to the fabrication of the active layer.

The $FAPbBr_3$ NCs were synthesized following our previously published procedure for $CsPbBr_3$ NCs[25] where, instead of Cesium acetate, we have employed Formamidine acetate as a precursor (see experimental section and Table S2 in the supporting information for the details on the volumes and precursors used). The main motivations to substitute Cs with FA is the necessity to purify the obtained material to produce films with reduced roughness. Nevertheless, LEDs based on FA-containing perovskites demonstrate comparable performance to Cs-containing ones, both when synthesized as NCs or nanostructured film.[26–28] In order to fabricate continuous solid films to be used in LEDs, we have developed a purification procedure that aims at reducing NC aggregation and removal of ligand excess without compromising the PLQY of the NCs. A schematic of the purification procedure employed is presented in **Figure 3a**. Importantly, $FAPbBr_3$ NCs demonstrate higher stability during purification compared to $CsPbBr_3$ synthesized with our procedure. In fact, in the case of $CsPbBr_3$ NCs, high PLQY is readily observed directly after synthesis[2,25] but further purification damages the NCs, thus showing a PLQY drop to 45 ± 5%. The obtained NC solution is initially centrifuged in order to transfer the NCs from the synthesis solvents mixture to anhydrous toluene. A post-synthesis treatment employing $PbBr_2$ complexes (see reference 25) is then applied to the NCs to enhance further the PLQY: pristine NCs show a PLQY of 35% directly after synthesis, the treatment enhances it up to 48%. After the treatment with $PbBr_2$, oleic acid is added in order to dissolve the NC aggregates that are observed in solution. Room-temperature synthetic approaches based on the use of relatively short-ligands are known to give rise to aggregation or coalescence during the NC crystallization.[2,25] Presence of aggregates/coalescence can be observed in TEM images of the as-synthesized NCs (see Figure S5), and it is further indicated by the turbid colour of the solution (Figure S6). Employing as-synthesized NCs for film preparation does not allow for the fabrication of working LEDs due to the high roughness of the obtained films (RMS roughness = 20.95 nm). Addition of few μl of oleic acid partially dissolves the aggregates (Figure 3b and photo in Figure S6). After this step the NC were centrifuged at slow speed to remove the large not-dissolved aggregates, and finally they were washed one last time with methyl acetate to remove excess ligands in solution.[4] Following this procedure we were able to fabricate spin-coated $FAPbBr_3$ NC films with an RMS roughness = 3.45 nm. Figure 3c shows the optical absorption and PL spectra of the $FAPbBr_3$ NCs in toluene solution. The NCs present a PL peak at 511 nm and FWHM = 22 nm. Spectral shape and PL peak position are unchanged compared to the not purified material. After the purification, the PLQY shows a 1.85 times increase up to 90 ± 9%. This enhancement is accompanied by a drastic change in the PL dynamics (Figure 3d) where the initial fast decay component reduces, thus increasing the PL lifetime from 2.66 ns to 4.27 ns (a summary of the PL decay fitting values is reported in Table S3). The PL lifetime after purification is 1.6 times longer than for the as-synthesized NCs. This increase is close to the one observed for the PLQY, thus indicating that the enhanced luminescence is correlated to a decrease in the non-radiative rate of the emitter (i.e. decrease of the fast component observed in the PL decay of the as-synthesized NCs). Films fabricated via spin-coating show a small red-shift of the PL peak (from 511 nm in solution to 516 nm in film), yet the PLQY is preserved, showing once again a value of 90 ± 9%. We have previously observed this surprisingly high PLQY in $CsPbBr_3$ NCs prepared with the very same synthetic method[3,25] and it can be tentatively assigned to hole trap passivation carried out by ambient oxygen in the spin-coated film.[29,30] These results show that the removal of aggregates in $FAPbBr_3$ NCs synthesized at room temperature and with short ligands is necessary to obtain highly luminescent material, in contrast with what observed for $CsPbBr_3$ NCs.

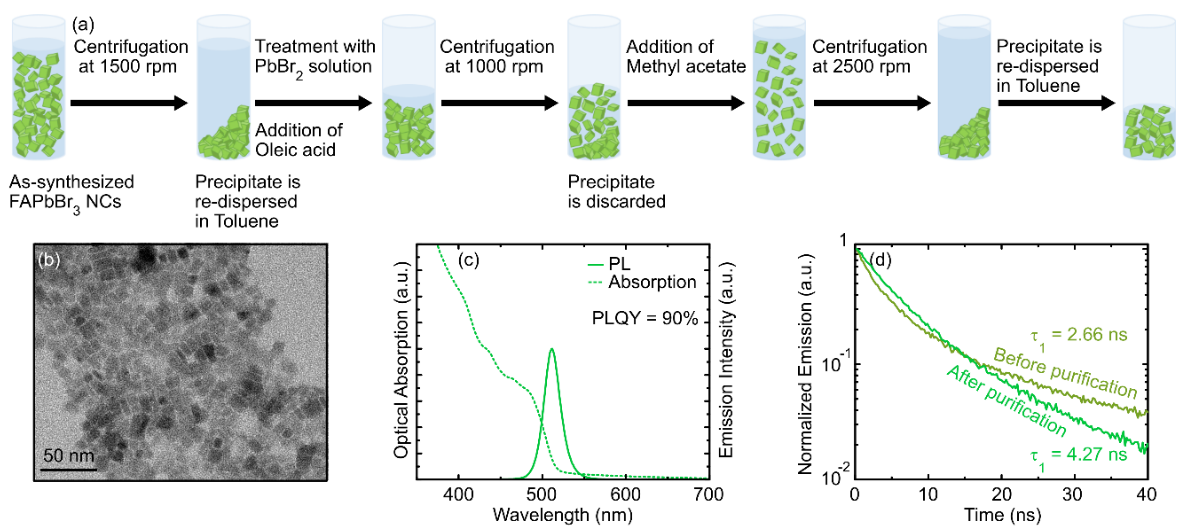

**Figure 3.** (a) Scheme of the purification procedure employed. The obtained $FAPbBr_3$ NCs were directly used for film fabrication via spin-coating after purification. (b) Transmission electron microscope micrograph of the $FAPbBr_3$ NCs after purification. (c) Optical absorption and steady-state PL spectrum of the purified $FAPbBr_3$ NCs in toluene solution. (c) Time-resolved PL of the $FAPbBr_3$ NCs in toluene solution before (dark green) and after (light green) purification, and magnification of the initial part of the PL decay (inset). The PL decays were fitted with a bi-exponential function and in the figure the dominant component ($\tau_1$) is reported.

Following the substitution of Cs with FA in our NCs, we have employed a cross-linking method to prevent dissolution of the NC film upon spin-coating of the hole-transport layer, thus removing the need to carry out an annealing step in order to insolubilize the NC layer. As reported by G. Li et al.,[9] exposure of a NC film to trimethylaluminum (TMA) vapours in an ALD system, followed by storage in air, it induces partial cross-linking of the surface ligands, enabling high film-retention rates (i.e. thickness reduction upon spin-rinsing with a solvent). We have used a similar procedure for our $FAPbBr_3$ NC film. Yet, due to the different synthetic approach and the relatively short length of the surface ligands in our synthesis, we obtained substantially different results compared to what observed for perovskite NCs prepared via hot-injection method. This cross-linking method is advantageous compared to ultra-high vacuum deposition of transport layers as it requires only few minutes to be completed and it is carried out at room temperature. In Figure S7 we report the impact of the ALD treatment (see experimental section and supporting information) on the luminescence of the $FAPbBr_3$ NC film spin-coated on glass. Directly after exposure of the film to TMA vapours, the PLQY of the film shows a drop from the original value of 90% to < 1% when 20 pulses of TMA vapours are applied. On the other hand, storage of the film in air at room-temperature for 24h induces a 201% increase of the PLQY. This effect is somewhat similar to what previously reported[9] but slower in comparison, as it is requires 24h of exposure to air. This can be caused by the shorter ligands used in our synthesis (octylamine/octanoic acid) which slows down the kinetic of the cross-linking reaction. Importantly, we do not observe an enhancement of the PLQY but only a recovery during storage in air. Following these results, we have then assessed the film-retention of the cross-linked $FAPbBr_3$ NC film. We observed that already 5 pulses of TMA vapours were sufficient to decrease the thickness loss to 10%, thus indicating that it is not necessary to sacrifice a substantial amount of luminescence to obtain an insoluble NC film. From these different parameters, we have found a trade-off condition between PLQY loss and film retention by using 3 pulses of TMA vapours, thus obtaining a film with a final PLQY of 46% and a film retention rate of 85%. Considering the performance of our LEDs, this indicates that a retention rate of 100% is not required to achieve high EQE and Luminance, and certain degree of NC film/hole-transport layer intermixing can be accepted. SEM imaging carried out on FAPbBr$_3$ NC film before and after cross-linking reveal no sizeable changes in film morphology and no cracks are formed following the procedure (see Figure S8 in the supporting information).

In conclusion, we have fabricated highly efficient LEDs employing solution-processed materials combined with a cross-linking procedure based on ALD, thus demonstrating that is possible to achieve high performance without the use of evaporated transport layers. The obtained devices outperform current perovskite NC LEDs based on solution-processed transport layers (EQE 6.04% vs 1.1%)[11] and LEDs based on FAPbBr$_3$ NCs (20.53 cd/A vs 13.02 cd/A).[31] Importantly, solution processed LEDs based on quantum dots of different composition or shape,[21,32] and conjugated polymers[33] are now demonstrating very competitive performance. We hope that our work will encourage further developments for perovskite NCs towards all-solution processed LEDs.

## AUTHOR INFORMATION

**Corresponding Author**

Gerasimos Konstantatos: Gerasimos.Konstantatos@icfo.eu
Francesco Di Stasio: Francesco.Distasio@icfo.eu

**Author Contributions**

‡⁺I. R. and Y. B. contributed equally to this work

**Funding Sources**

The authors acknowledge financial support from the European Research Council (ERC) under the European Union's Horizon 2020 research and innovation programme (grant agreement no. 725165), the Spanish Ministry of Economy and Competitiveness (MINECO), and the "Fondo Europeo de Desarrollo Regional" (FEDER) through grant TEC2017-88655-R and the program redes de excellencia TFE by MINECO. The authors also acknowledge financial support from Fundacio Privada Cellex, the program CERCA and from the Spanish Ministry of Economy and Competitiveness, through the "Severo Ochoa" Programme for Centres of Excellence in R&D (SEV-2015-0522). F. Di Stasio and S. Christodoulou acknowledge support from two Marie Curie Standard European Fellowships ("NANOPTO", H2020-MSCA-IF-2015-703018 and "NAROBAND", H2020-MSCA-IF-2016-750600). I. Ramiro acknowledges support from the Ministerio de Economía, Industria y Competitividad of Spain via a Juan de la Cierva fellowship.

## REFERENCES

(1) Akkerman, Q. A.; Rainò, G.; Kovalenko, M. V; Manna, L. Genesis, Challenges and Opportunities for Colloidal Lead Halide Perovskite Nanocrystals. *Nat. Mater.* **2018**, *17*, 394–405.

(2) Akkerman, Q. A.; Gandini, M.; Di Stasio, F.; Rastogi, P.; Palazon, F.; Bertoni, G.; Ball, J. M.; Prato, M.; Petrozza, A.; Manna, L. Strongly Emissive Perovskite Nanocrystal Inks for High-Voltage Solar Cells. *Nat. Energy* **2017**, *2*, 16194.

(3) Christodoulou, S.; Di Stasio, F.; Pradhan, S.; Stavrinadis, A.; Konstantatos, G. High-Open-Circuit-Voltage Solar Cells Based on Bright Mixed-Halide CsPbBrI$_2$ Perovskite Nanocrystals Synthesized under Ambient Air Conditions. *J. Phys. Chem. C* **2018**, *122*, 7621-7626.

(4) Swarnkar, A.; Marshall, A. R.; Sanehira, E. M.; Chernomordik, B. D.; Moore, D. T.; Christians, J. A.; Chakrabarti, T.; Luther, J. M. Quantum Dot–induced Phase Stabilization of α-CsPbI$_3$ Perovskite for High-Efficiency Photovoltaics. *Science* **2016**, *354*, 92-95..

(5) Meinardi, F.; Akkerman, Q. A.; Bruni, F.; Park, S.; Mauri, M.; Dang, Z.; Manna, L.; Brovelli, S. Doped Halide Perovskite Nanocrystals for Reabsorption-Free Luminescent Solar Concentrators. *ACS Energy Lett.* **2017**, *2*, 2368–2377.

(6) Nedelcu, G.; Protesescu, L.; Yakunin, S.; Bodnarchuk, M. I.; Grotevent, M. J.; Kovalenko, M. V. Fast Anion-Exchange in Highly Luminescent Nanocrystals of Cesium Lead Halide Perovskites (CsPbX3, X = Cl, Br, I). *Nano*

*Lett.* **2015**, *15*, 5635–5640.

(7) Pan, J.; Sarmah, S. P.; Murali, B.; Dursun, I.; Peng, W.; Parida, M. R.; Liu, J.; Sinatra, L.; Alyami, N.; Zhao, C.; Alarousu, E.; Ng, T. K.; Ooi, B. S.; Bakr, O. M.; Mohammed, O. F. Air-Stable Surface-Passivated Perovskite Quantum Dots for Ultra-Robust, Single- and Two-Photon-Induced Amplified Spontaneous Emission. *J. Phys. Chem. Lett.* **2015**, *6*, 5027–5033.

(8) Palazon, F.; Di Stasio, F.; Akkerman, Q. A.; Krahne, R.; Prato, M.; Manna, L. Polymer-Free Films of Inorganic Halide Perovskite Nanocrystals as UV-to-White Color-Conversion Layers in LEDs. *Chem. Mater.* **2016**, *28*, 2902-2906.

(9) Li, G.; Rivarola, F. W. R.; Davis, N. J. L. K.; Bai, S.; Jellicoe, T. C.; de la Peña, F.; Hou, S.; Ducati, C.; Gao, F.; Friend, R. H.; Greenham, N. C.; Tan, Z.-K. Highly Efficient Perovskite Nanocrystal Light-Emitting Diodes Enabled by a Universal Crosslinking Method. *Adv. Mater.* **2016**, *28*, 3528–3534.

(10) Chiba, T.; Hoshi, K.; Pu, Y.-J.; Takeda, Y.; Hayashi, Y.; Ohisa, S.; Kawata, S.; Kido, J. High-Efficiency Perovskite Quantum-Dot Light-Emitting Devices by Effective Washing Process and Interfacial Energy Level Alignment. *ACS Appl. Mater. Interfaces* **2017**, *9*, 18054–18060.

(11) Shamsi, J.; Rastogi, P.; Caligiuri, V.; Abdelhady, A. L.; Spirito, D.; Manna, L.; Krahne, R. Bright-Emitting Perovskite Films by Large-Scale Synthesis and Photoinduced Solid-State Transformation of CsPbBr3 Nanoplatelets. *ACS Nano* **2017**, *11*, 10206–10213.

(12) Veldhuis, S. A.; Ng, Y. F.; Ahmad, R.; Bruno, A.; Jamaludin, N. F.; Damodaran, B.; Mathews, N.; Mhaisalkar, S. G. Crown Ethers Enable Room-Temperature Synthesis of CsPbBr3 Quantum Dots for Light-Emitting Diodes. *ACS Energy Lett.* **2018**, *3*, 526–531.

(13) Kim, Y.; Yassitepe, E.; Voznyy, O.; Comin, R.; Walters, G.; Gong, X.; Kanjanaboos, P.; Nogueira, A. F.; Sargent, E. H. Efficient Luminescence from Perovskite Quantum Dot Solids. *ACS Appl. Mater. Interfaces* **2015**, *7*, 25007–25013.

(14) Protesescu, L.; Yakunin, S.; Bodnarchuk, M. I.; Krieg, F.; Caputo, R.; Hendon, C. H.; Yang, R. X.; Walsh, A.; Kovalenko, M. V. Nanocrystals of Cesium Lead Halide Perovskites (CsPbX3, X = Cl, Br, and I): Novel Optoelectronic Materials Showing Bright Emission with Wide Color Gamut. *Nano Lett.* **2015**, *15*, 3692–3696.

(15) De Roo, J.; Ibáñez, M.; Geiregat, P.; Nedelcu, G.; Walravens, W.; Maes, J.; Martins, J. C.; Van Driessche, I.; Kovalenko, M. V; Hens, Z. Highly Dynamic Ligand Binding and Light Absorption Coefficient of Cesium Lead Bromide Perovskite Nanocrystals. *ACS Nano* **2016**, *10*, 2071–2081.

(16) Yan, F.; Xing, J.; Xing, G.; Quan, L.; Tan, S. T.; Zhao, J.; Su, R.; Zhang, L.; Chen, S.; Zhao, Y.; Huan, A.; Sargent, E. H.; Xiong, Q.; Demir, H. V. Highly Efficient Visible Colloidal Lead-Halide Perovskite Nanocrystal Light-Emitting Diodes. *Nano Lett.* **2018**, *18*, 3157–3164.

(17) Hung-Chia, W.; Zhen, B.; Hsin-Yu, T.; An-Cih, T.; Ru-Shi, L. Perovskite Quantum Dots and Their Application in Light-Emitting Diodes. *Small* **2017**, *14*, 1702433.

(18) Palazon, F.; Di Stasio, F.; Lauciello, S.; Krahne, R.; Prato, M.; Manna, L. Evolution of CsPbBr3 nanocrystals upon Post-Synthesis Annealing under an Inert Atmosphere. *J. Mater. Chem. C* **2016**, *4*, 9179-9182.

(19) K., R. A.; F., P. G. de A.; Alexandros, S.; Tania, L.; Maria, B.; L., D. S.; Gerasimos, K. Remote Trap Passivation in Colloidal Quantum Dot Bulk Nano-heterojunctions and Its Effect in Solution-Processed Solar Cells. *Adv. Mater.* **2014**, *26*, 4741–4747.

(20) Rastogi, P.; Palazon, F.; Prato, M.; Di Stasio, F.; Krahne, R. Enhancing the Performance of CdSe/CdS Dot-in-Rod Light-Emitting Diodes via Surface Ligand Modification. *ACS Appl. Mater. Interfaces* **2018**, *10*, 5665-5672.

(21) Dai, X.; Zhang, Z.; Jin, Y.; Niu, Y.; Cao, H.; Liang, X.; Chen, L.; Wang, J.; Peng, X. Solution-Processed, High-Performance Light-Emitting Diodes Based on Quantum Dots. *Nature* **2014**, *515*, 96–99.

(22) Li, J.; Shan, X.; Bade, S. G. R.; Geske, T.; Jiang, Q.; Yang, X.; Yu, Z. Single-Layer Halide Perovskite Light-Emitting Diodes with Sub-Band Gap Turn-On Voltage and High Brightness. *J. Phys. Chem. Lett.* **2016**, *7*, 4059–4066.

(23) Schubert, E. F. *Light-Emitting Diodes*, 2nd ed.; Cambridge University Press: Cambridge, 2006.

(24) Greenham, N. C.; Friend, R. H.; Bradley, D. D. C. Angular Dependence of the Emission from a Conjugated Polymer Light-Emitting Diode: Implications for Efficiency Calculations. *Adv. Mater.* **1994**, *6*, 491–494.

(25) Di Stasio, F.; Christodoulou, S.; Huo, N.; Konstantatos, G. Near-Unity Photoluminescence Quantum Yield in CsPbBr3 Nanocrystal Solid-State Films via Post-Synthesis Treatment with Lead Bromide. *Chem. Mater.* **2017**, *29*, 7663–7667.

(26) Chin, X. Y.; Perumal, A.; Bruno, A.; Yantara, N.; Veldhuis, S. A.; Martínez-Sarti, L.; Chandran, B.; Chirvony, V.; Lo, A. S.-Z.; So, J.; Soci, C.; Grätzel, M.; Bolink, H. J.; Mathews, N.; Mhaisalkar, S. G. Self-Assembled Hierarchical Nanostructured Perovskites Enable Highly Efficient LEDs via an Energy Cascade. *Energy Environ. Sci.* **2018**, *11*, 1770–1778.

(27) Zhang, X.; Wang, W.; Xu, B.; Liu, H.; Shi, H.; Dai, H.; Zhang, X.; Chen, S.; Wang, K.; Sun, X. W. Less-Lead Control toward Highly Efficient Formamidinium-Based Perovskite Light-Emitting Diodes. *ACS Appl. Mater. Interfaces* **2018**, *10*, 24242–24248.

(28) Yang, X.; Zhang, X.; Deng, J.; Chu, Z.; Jiang, Q.; Meng, J.; Wang, P.; Zhang, L.; Yin, Z.; You, J. Efficient Green Light-Emitting Diodes Based on Quasi-Two-Dimensional Composition and Phase Engineered Perovskite with Surface Passivation. *Nat. Commun.* **2018**, *9*, 570.

(29) Lorenzon, M.; Sortino, L.; Akkerman, Q.; Accornero, S.; Pedrini, J.; Prato, M.; Pinchetti, V.; Meinardi, F.; Manna, L.; Brovelli, S. Role of Nonradiative Defects and Environmental Oxygen on Exciton Recombination Processes in $CsPbBr_3$ Perovskite Nanocrystals. *Nano Lett.* **2017**, *17*, 3844–3853.

(30) Meggiolaro, D.; Mosconi, E.; De Angelis, F. Mechanism of Reversible Trap Passivation by Molecular Oxygen in Lead-Halide Perovskites. *ACS Energy Lett.* **2017**, *2*, 2794–2798.

(31) Kumar, S.; Jagielski, J.; Kallikounis, N.; Kim, Y.-H.; Wolf, C.; Jenny, F.; Tian, T.; Hofer, C. J.; Chiu, Y.-C.; Stark, W. J.; Lee, T.-W.; Shih, C.-J. Ultrapure Green Light-Emitting Diodes Using Two-Dimensional Formamidinium Perovskites: Achieving Recommendation 2020 Color Coordinates. *Nano Lett.* **2017**, *17*, 5277–5284.

(32) Giovanella, U.; Pasini, M.; Lorenzon, M.; Galeotti, F.; Lucchi, C.; Meinardi, F.; Luzzati, S.; Dubertret, B.; Brovelli, S. Efficient Solution-Processed Nanoplatelet-Based Light-Emitting Diodes with High Operational Stability in Air. *Nano Lett.* **2018**, *18*, 3441–3448.

(33) Di, D.; Romanov, A. S.; Yang, L.; Richter, J. M.; Rivett, J. P. H.; Jones, S.; Thomas, T. H.; Abdi Jalebi, M.; Friend, R. H.; Linnolahti, M.; Bochmann, M.; Credgington, D. High-Performance Light-Emitting Diodes Based on Carbene-Metal-Amides. *Science* **2017**, *356*, 159-163.

# Supporting information

# High-Efficiency Light-Emitting Diodes Based on Formamidinium Lead Bromide Nanocrystals and solution processed transport layers

Francesco Di Stasio[†], Iñigo Ramiro[†], Yu Bi[†], Sotirios Christodoulou[†], Alexandros Stavrinadis[†] and Gerasimos Konstantatos[†,*]

[†]ICFO-Institut de Ciencies Fotoniques, The Barcelona Institute of Science and Technology, 08860 Castelldefels (Barcelona), Spain
[*]ICREA—Institució Catalana de Recerca i Estudis Avançats, Passeig Lluís Companys 23, 08010

### Chemicals:

Lead(II) bromide ($PbBr_2$, 99.999% trace metals basis), Formamidine acetate (FAAc, 99%), Octylamine (OcAm, 99%), Butylamine (ButAm, 99.5%) Octanoic acid (OcAc, 98%), Propionic acid (PrAc, > 99.5%), Oleic acid (99%), anhydrous Toluene (TOL, 99.8%), Methyl acetate (>98%), anhydrous Chlorobenzene (99.8%) and anhydrous 1,4 dioxane (99.8%) were purchased from Sigma-Aldrich. 1-propanol (PrOH, Pharmpur®) and n-Hexane (Hex, 99%) were purchased from Sharlab. All chemicals were used without any further purification.

### Experimental:

*Perovskite nanocrystals synthesis and purification*: The NCs were synthesized using the previously published procedure in ref. 25 of the main text with a modification: instead of using Cs-acetate dissolved in 1-propanol, we have employed a 1-propanol formamidine acetate solution (18 mg/ml). The as-synthesized NCs were then centrifuged at 1500 rpm for 2 minutes and re-dispersed in 20 ml of anhydrous toluene. A post-synthesis treatment with $PbBr_2$ (185 mg/ml) dissolved in 1:1:1 Propionic Acid:Butylamine:Hexane in volume was carried out to enhance the PLQY (see ref. 25, 12 μl for a typical synthesis). After the treatment, oleic acid (35 μl for a typical synthesis) was added to the solution to dissolve the NC aggregates observed under TEM. The solution was then centrifuged at 1000 rpm, the supernatant was collected and transferred into a new vial. An equal volume of anhydrous Methyl acetate (20 ml in a typical reaction) was added to the solution and centrifuged once more at 2500 rpm for 5 minutes, the supernatant was discarded and the precipitate re-dissolved in 500 μl anhydrous toluene. Finally, the obtained NC solution was ready for film preparation. The entire procedure was carried out in air and at room temperature.

*Transmission Electron Microscope:* Imaging was carried out with a JEOL JEM-2100 LaB6 transmission electron microscope, operating at 200 kV. The spectrometer is an Oxford Instruments INCA x-sight, with a Si(Li) detector. Samples for TEM characterization were prepared by drop-casting diluted NC solutions onto 300-mesh carbon-coated copper grids.

*Ultraviolet photoelectron spectroscopy:* UPS measurements were performed with a SPECS PHOIBOS 150 hemispherical analyzer (SPECS GmbH, Berlin, Germany) in ultra-high vacuum conditions ($10^{-10}$ mbar) using a monochromatic HeI UV source (21.2 eV) as excitation.

*Device cross section:* The cross sectional images of the device were obtained using a Zeiss Augira cross-beam workstation. A layer of platinum (Pt, around 200 nm) was deposited via gas injection system while the cross-section cut was made with a gallium focus ion beam (Ga-FIB). The SEM image was obtained using a voltage of 5kV and aperture size of 30μm with an Inlens detector.

*Optical characterization:* Optical absorption spectra were collected using a Varian Cary-5000 UV-Vis-NIR spectrophotometer. Photoluminescence (PL) measurements were performed using a Horiba Jobin Yvon iHR550 Fluorolog system coupled to a Horiba TBX-04 photomultiplier tube, a calibrated Quanta-phi integrating sphere and a FluoroHub time-correlated single photon counting card. All steady-state PL spectra were corrected for the system response function and were collected using a Xenon-lamp coupled with a monocromator as excitation source. For time-resolved PL measurements a pulsed laser diode was employed (Horiba Nanoled, $\lambda$ = 405 nm, pulse full-width-half-maximum of 50 ps, fluence $\approx$ 1 nJ/cm$^2$). All photoluminescence quantum yield (PLQY) measurements were carried out in the integrating sphere ($\lambda_{exc}$ = 380 nm, $\Delta\lambda$ = 5 nm, power density = 1 mW/cm$^2$). FAPbBr$_3$ NCs solutions for PLQY were prepared in quartz cuvettes and diluted to 0.1 optical density at the excitation wavelength and the measurements were carried out in a 4π configuration. For PLQY on films, the measurements were carried out in a 2π configuration, and all films were prepared via spin-coating at 2000 rpm on soda-lime glass substrates (area of 1 cm$^2$).

*Device fabrication:* Glass slides with patterned ITO were used as substrates and they were cleaned initially with soapy water, rinsed then immersed in acetone for 30 minutes, and finally rinsed with isopropanol. The electron transport layer was deposited via spin-coating in air at 3000 rpm for 30 sec of a ZnO nanoparticle (40 mg/ml) chloroform solution (ZnO nanoparticles were previously synthesized using the procedure in ref. 19 of the main text). The spin-coating step was repeated a second time to obtain the desired thickness (80 nm) followed by annealing in air at 260C for 30 minutes. The FAPbBr$_3$ NC layer was spin-coated in air at 2000 rpm for 20 sec (solution concentration $\approx$ 120 mg/ml, thickness = 95 nm); the film RMS was measured with a Dektak Profilometer. Afterward, the glass/ITO/ZnO/FAPbBr$_3$ NC layer substrates were transferred in an ALD system (GEMStar XT Thermal, Arradiance inc.) inside a Nitrogen filled glove-box where they were exposed to 3 pulses of trimethylaluminium to carry out the ligand cross-linking step. The substrates were then left in air for 24h and after transferred into a nitrogen filled glove-box where the hole-transport layer was spin-coated using a 10 mg/ml poly vinyl carbazole (PVK) solution in 1,4-dioxane and a 10 mg/ml Poly[N,N'-bis(4-butylphenyl)-N,N'-bisphenylbenzidine] (PTPD) in chlorobenzene; both polymers were spincoated at 2000 rpm for 30 sec (total thickness: 95 nm). The 50 nm Platinum electrode was deposited using a shadow mask and a RC magnetron sputtering system (AJA Orion 8 HV) at room temperature.

*Device characterization:* The current-voltage-luminance characteristics were measured using a Keithley 2636A source-measure unit coupled to a calibrated PDA 100A Si switchable gain detector from Thorlabs. The system was controlled via a LabView interface. The output of the Si detector was converted into power (photon flux) using the responsivity of the detector. The EQE was calculated as the ratio of the photon flux and the driving current of the device. The EL spectra of the devices were collected using the Horiba Jobin Yvon iHR550 Fluorolog system previously described while the bias was manually applied with a Keithley 2636A source-measure unit.

## Ultraviolet Photoelectron spectroscopy:

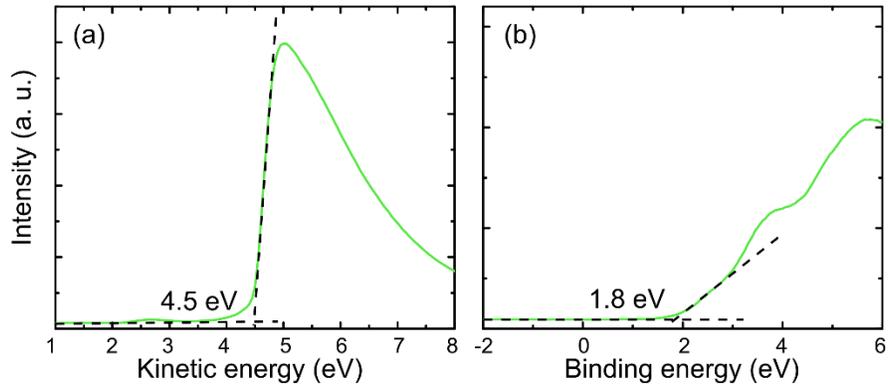

**Figure S1** UPS spectrum of a FAPbBr$_3$ NC film prepared via spin-coating on ITO glass. (a) Secondary electron emission as a function of the kinetic energy of the photoelectrons. From the onset of the photoemission a work function of (4.5±0.2) eV can be estimated for the FAPbBr$_3$ NC film. (b) Valence band maximum (VBM) relative to the Fermi level. From the VBM onset a value of 1.8±0.2 eV is extracted, thus obtaining a position of the VBM with respect to the vacuum level of -6.3±0.2 eV.

## Current density-luminance-voltage curves of the "champion" device in logarithmic scale:

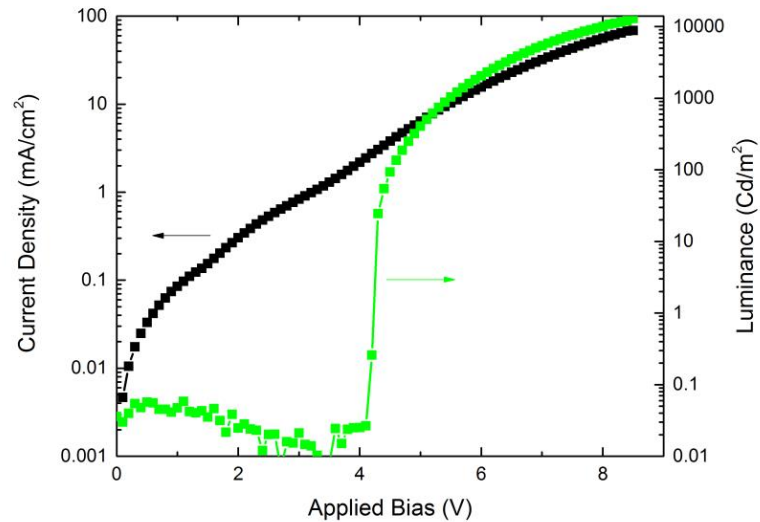

**Figure S2** JVL curves of the "champion" device plotted in logarithmic scale for clarity.

CIE 1931 colour coordinates values:

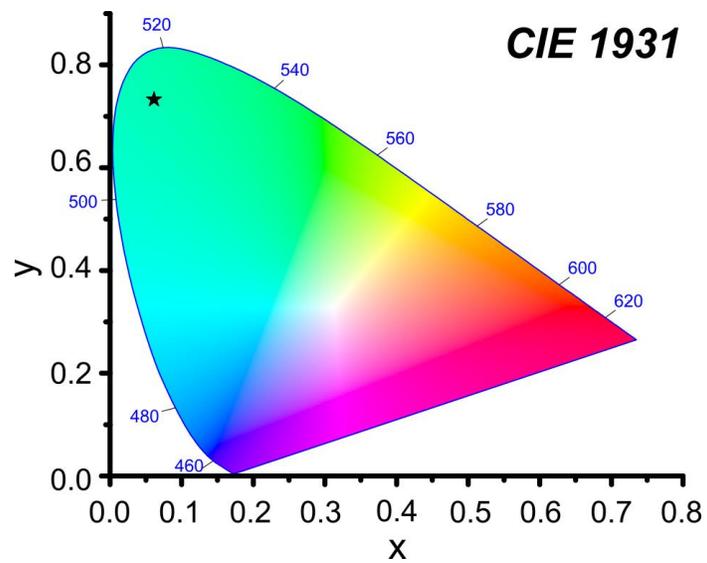

**Figure S3** CIE 1931 diagram for FAPbBr$_3$ NC LEDs. The obtained color coordinates are x = 0.0613 and y = 0.732

## LED performance VS FAPbBr$_3$ NC film thickness:

Table S1 key LED performance parameters variation upon FAPbBr$_3$ NC film thickness increase

| Thickness (nm) | V$_{ON}$ (V) | EQE$_{MAX}$ (%) |
|---|---|---|
| 120 | 4.7 | 5.31 |
| 95 | 4.1 | 6.04 |
| 60 | 3.2 | 3.72 |
| 40 | 2.9 | 2.45 |

## JVL curve acquired after first scan:

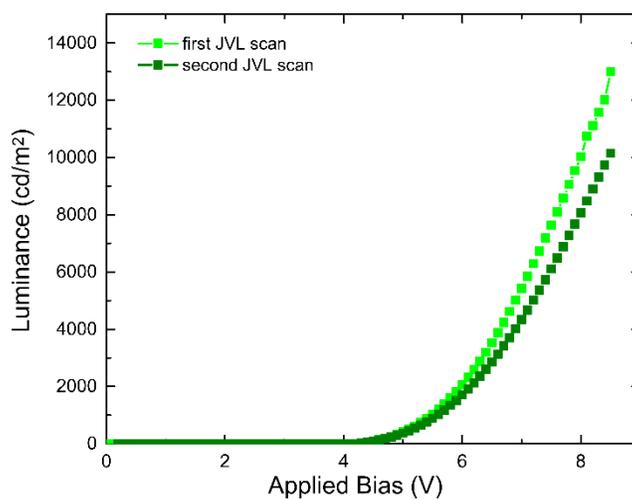

Figure S4 Device stability remains very limited despite the good overall performance of the LEDs. A JVL scan carried out directly after the intial one already present a decrease in Luminance from 12998 cd/m² to 10147 cd/m²

TEM micrograph of the FAPbBr$_3$ NCs before and after purification:

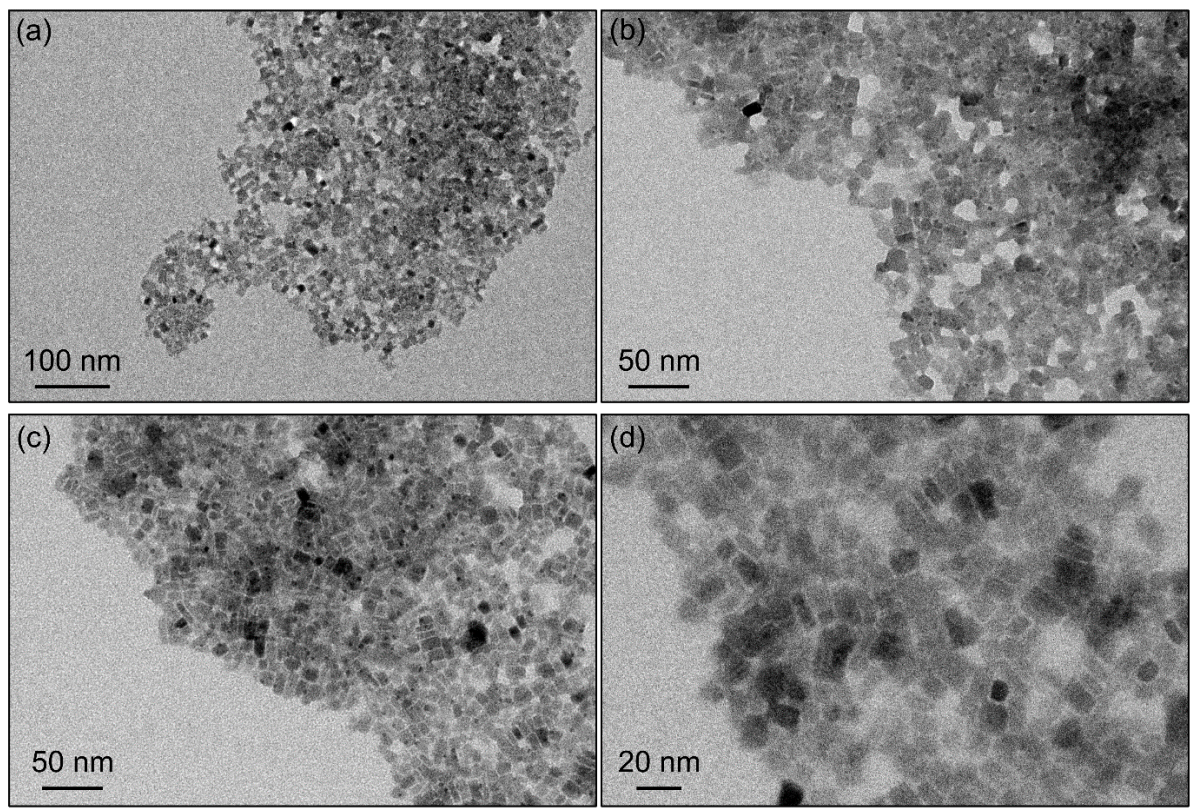

**Figure S5** TEM micrographs at different magnifications of the FAPbBr$_3$ NCs before (a, b) and after (c, d) the addition of oleic acid during purification. The as-synthesized NCs appear aggregated before addition of the oleic acid, as observed in similar synthetic methods emplyoing short ligands and carried out at room temperature.

## Effect of the addition of oleic acid to the FAPbBr$_3$ NC solution:

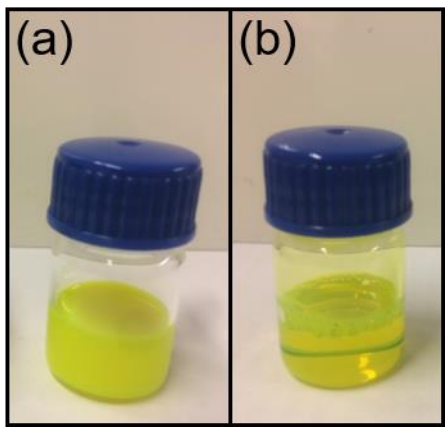

**Figure S6** Photo of the FAPbBr$_3$ NC solution before (a) and after (b) the oleic acid addition. The NC solution turns from turbid to clear.

## Total volumes used in a typical reaction:

**Table S2** Total solvent volumes and precursors weigths used in a typical reaction of FAPbBr$_3$ NCs. 6 ml of the FA precursor solution (18 mg/ml) was first added to the Hexane and 1-propanol mixture. Afterward, 7ml of the PbBr$_2$ precursor solution (185 mg/ml) was quickly injected into the flask. (see reference 25 of the main text)

| Chemicals | Reaction flask | Formamidinium precursor solution | PbBr$_2$ precursor solution |
|---|---|---|---|
| Hexane | 48 ml | / | / |
| 1-propanol | 18 ml | 6 ml | 2.33 ml |
| Octanoic acid | / | / | 2.33 ml |
| Octylamine | / | / | 2.33 ml |
| Formamidine acetate | / | 108 mg | / |
| PbBr$_2$ | / | / | 1295 mg |

## PL decays fitting parameters:

**Table S3** PL decays were fitted with a two-exponential function: $I = I_0 + I_1 e^{(-t/\tau_1)} + I_2 e^{(-t/\tau_2)}$ using the following parameters

| Sample | I$_1$ | $\tau_1$ (ns) | I$_2$ | $\tau_2$ (ns) |
|---|---|---|---|---|
| FAPbBr$_3$ NCs before purification | 0.71 | 2.66 | 0.29 | 14.1 |
| FAPbBr$_3$ NCs after purification | 0.75 | 4.27 | 0.25 | 13.85 |

## Cross-linking of the FAPbBr₃ NC film with TMA vapours:

The cross-linking of the film was performed in a GEMStar XT Thermal ALD system. High-purity TMA was purchased from STREM Chemicals Inc. The process was carried out close to ambient temperature (29-33 °C). Before the process, the reaction chamber was pumped down and subsequently filled with pure nitrogen up to a pressure of typically 1.3-1.4 mbar. The TMA manifold was maintained at 150 °C during gas supply. TMA was applied in 9-ms pulses (longer pulses were found to quench the film luminescence) at a partial pressure of 0.04 mbar. The dosage of TMA in different samples depends on the number of TMA pulses applied.

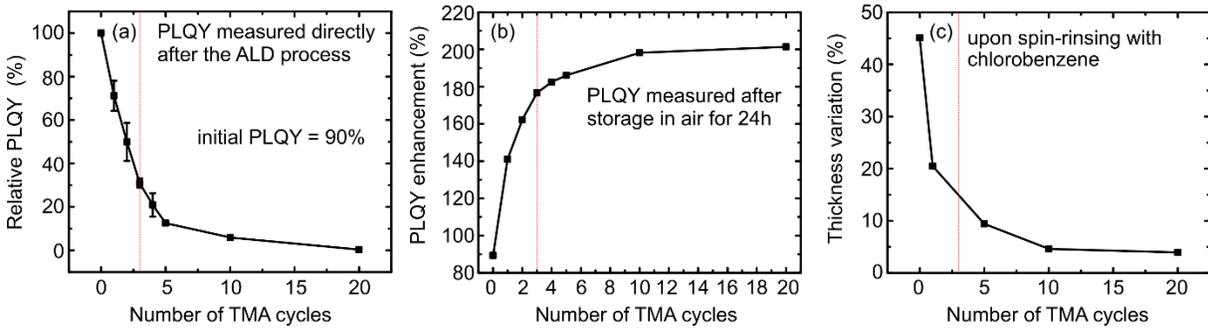

**Figure S7** Effect of the TMA vapours treatment on the FAPbBr₃ NC film. (a) PLQY variation after the treatment vs number of TMA vapours cycles. (b) PLQY recovery and enhancement after exposure to air for 24h vs number of TMA vapours cycles previously applied. (c) NC Film thickness reduction upon spin-rinsing with chlorobenzene vs number of TMA vapours cycles. The FAPbBr₃ NC film used in the LED was exposed to 3 pulses of TMA (highlighted with dotted red line in all panels) thus obtaining a film with a retention rate of 85% and a final PLQY of 46%.

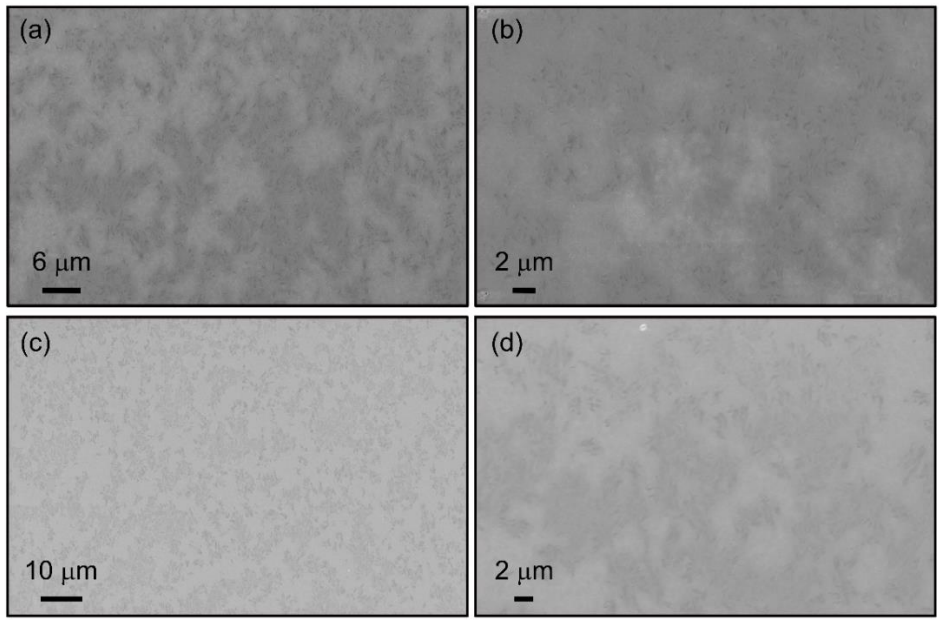

**Figure S8** SEM images of the FAPbBr₃ NC film before (a, b) and after (c, d) exposure to TMA vapours. No cracks or sizeable changes in film morphology are observed.